\documentclass{book}
\usepackage{amsmath}
\usepackage{amstext}
\usepackage{amssymb}
\usepackage{array}
\usepackage{longtable}
\usepackage{science}
\usepackage{graphicx}
\usepackage{makeidx}
\usepackage{epstopdf}
\usepackage{rotating}
\usepackage{lscape}
\usepackage[marginal]{footmisc}
\usepackage{booktabs}
\usepackage{threeparttable}

\makeindex
\begin{document}

\chapter{Cognitive State Analysis, Understanding, and Decoding from the Perspective of Brain Connectivity}

\textbf{\large{Junhua Li\footnotemark{}\footnotetext{Correspondence should be addressed to Junhua Li (juhalee.bcmi@gmail.com)}, Anastasios Bezerianos, and Nitish Thakor}}\\\\

\section*{Abstract}
 Cognitive states are involving in our daily life, which motivates us to explore them and understand them by a vast variety of perspectives. Among these perspectives, brain connectivity is increasingly receiving attention in recent years. It is the right time to summarize the past achievements, serving as a cornerstone for the upcoming progress in the field. In this chapter, the definition of the cognitive state is first given and the cognitive states that are frequently investigated are then outlined. The introduction of the methods for estimating connectivity strength and graph theoretical metrics is followed. Subsequently, each cognitive state is separately described and the progress in cognitive state investigation is summarized, including analysis, understanding, and decoding. We concentrate on the literature ascertaining macro-scale representations of cognitive states from the perspective of brain connectivity and give an overview of achievements related to cognitive states to date, especially within the past ten years. The discussions and future prospects are stated at the end of the chapter.
 
      \begin{figure}[!bh]
 \centering
 \includegraphics[width=0.9\textwidth]{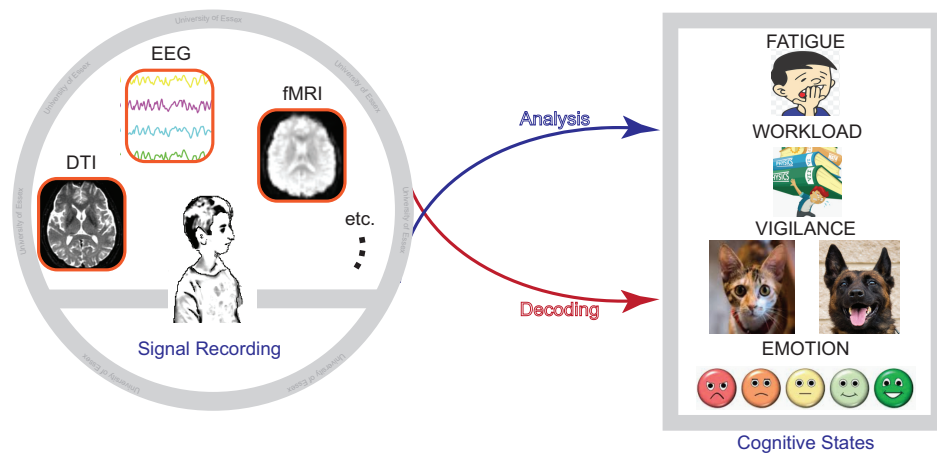}
 \caption{Overview of analysis and decoding of cognitive states using a variety of signals including time series modality and image modality. Different signal modalities have their own pros and cons. What modality is adopted depends on the need of an experiment and the purposes of a study. They can be either individually adopted  or jointly adopted. This figure incorporates some parts of images from the Google Images with labelling the permission of reuse and some parts of images from \cite{li2010bilateral, li2014deep} with permissions given by Elsevier and Springer.}
 \label{figure1}
 \end{figure}

\section{Introduction}
Cognitive state is defined as the state of a person's cognitive processes in a dictionary. It reflects the underlying mental action and process related to a wide range of cognitions such as memory, attention, evaluation, reasoning, problem-solving, comprehension, and language organization. Therefore, cognitive state assessment endows to probe into cognition and be aware of the brain state. Questionnaire-based assessment is a quick and easy way to gauge cognitive state with the advantages of low cost and unlimited accessibility. This method is useful and effective to assess quite stable cognitive statuses originated from pathological conditions such as dementia because the cognitive ability is almost not affected by subjective factors. For instance, the questionnaires of Mini-Mental State Examination (MMSE) \cite{Folstein1975} and Montreal cognitive assessment (MoCA) \cite{Nasreddine2005} are prevalently utilized to assess cognitive capability and to detect cognitive decline. However, questionnaire-based assessment is not very applicable to measure cognitive states which could vary in a short period. Another drawback of the use of questionnaire-based assessment is the incapability of a real-time evaluation. It depends on the recall of past engagement and is largely affected by subjective factors. In contrast, the assessment based on neurophysiological signals provides an objective evaluation of cognitive states and can perform a real-time evaluation, but this comes at the expense of high cost and is restricted by the availability of measuring equipment and a trained operator. To date, a variety of neurophysiological signals have been being utilized to assess diverse cognitive states. The signals encompass time series modality (e.g., electroencephalogram (EEG), magnetoencephalogram (MEG)) and image modality (e.g., functional magnetic resonance imaging (fMRI), diffusion tensor imaging (DTI)). These signals are analysed (analysis) or decoded (classification) to understand human cognitive states (see Fig. \ref{figure1}). For the analysis, the objective is to find intrinsic characteristics that are associated with a given cognitive state or to reveal representative differences between different levels of a cognitive state or between different groups (e.g., patient group versus healthy group). The resultant findings from the analysis could inform feature extraction in the classification so that informative and discriminative features can be extracted to facilitate the classification.   

 \begin{figure}[bht]
 \centering
 \includegraphics[width=0.9\textwidth]{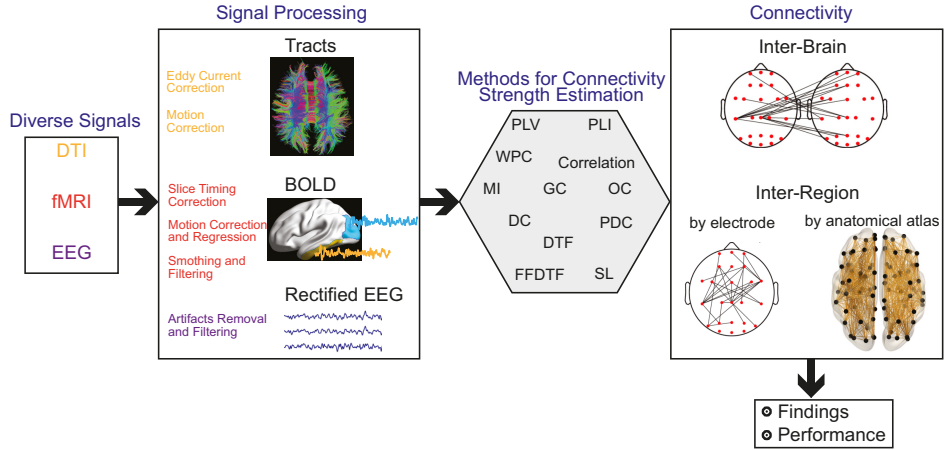}
 \caption{Illustration of the main flowchart for the analysis and classification of cognitive states. The signals recorded from the human are processed to remove interferences and retain cognition-related information. The processing steps and methods vary for different modalities. The critical processing steps for three modalities are delineated in the panel of signal processing. Connective strengths between brains (inter-brain) or between brain regions (inter-region) are estimated by an approach listed in the hexagon to construct a connectivity network. Subsequently, analysis and classification are performed based on the connectivity network. The findings revealing brain connectivity representations related to cognitive states are presented in the analysis studies while the classification accuracies are reported in the classification studies. The approaches for connectivity estimation are Phase Locking Value, PLV; Phase Lag Index, PLI; Wavelet Phase Coherence, WPC; Pearson Correlation and Cross Correlation, Correlation; Mutual Information, MI; Granger Causality, GC; Ordinary Coherence, OC; Directed Coherence, DC; Partial Directed Coherence, PDC; Directed Transfer Function, DTF; Full Frequency Directed Transfer Function, FFDTF; Synchronization Likelihood, SL.  
 This figure incorporates some parts of images from \cite{Harvy2019, sun2017asymmetry} with the reuse permission under an open licence.}
 \label{figure2}
 \end{figure} 
 
No matter a study is for analysing and/or decoding cognitive states, a signal measured from a specific location can be considered the proxy of brain activity of that location. In general, signals are recorded from the human by placing measuring electrodes or sensors at the targeted locations or scanning a particular body part. The recorded signals should be first processed to remove interferential components (i.e., artifacts). For different signal modalities, the processing of artifact removal varies. The general and critical processing steps and methods for three frequently-adopted modalities were shown in the panel of signal processing in Fig. \ref{figure2}. After the artifact removal, individual regions are separately explored to investigate cognitive states. This manner has been adopted by numerous studies. In this way, brain regions are isolated when exploring characteristics relevant to the cognitive state. However, brain regions cooperate together to perform a task while each region might dominantly contribute to a particular function required in the task. Therefore, the exploration of interregional interactions in the brain is necessary to account for the cooperation. A vast range of approaches can be utilized to estimate the interregional interactions in terms of connectivity strength (see the approaches enclosed in the hexagon in Fig. \ref{figure2}). Subsequently, cognitive states are explored based on the connectivity. To date, researchers have accomplished some achievements and made progress in the investigation of cognitive states from the perspective of brain connectivity. We summarize these achievements in this chapter to give an overview of the progress. Due to diverse cognitive states and ambiguous boundaries among cognitive states, we, in this chapter, focus on mental fatigue, mental workload, vigilance, and emotion, which have been relatively frequently ascertained and reported in the literature. The chapter was drafted in March 2019 and revised by incorporating some selected papers afterwards in August 2020. The remainder of the chapter is organized as follows. We describe the methods of estimating connectivity strength and graph theoretical metrics in the next section (Sect. \ref{1.2}). This is followed by the sections stating the details of achievements in mental fatigue (Sect. \ref{MentalFatigue}), mental workload (Sec. \ref{MentalWorkload}), vigilance (Sect. \ref{Vigilance}), and emotion (Sect. \ref{Emotion}). In Sect. \ref{Toolbox}, toolboxes used for brain connectivity analysis and classification are introduced. After that, thoughts and further directions are drawn in Sect. \ref{TFD}. Finally, a conclusion is given in Sect. \ref{Conclusion}.      

\section{Methods for Connectivity} 
\label{1.2}
Connectivity is a representation of relationships between brain regions or between channels. The extent of interactions between regions or between channels is quantified by the methods estimating connectivity strength and is then represented as a connectivity matrix. Subsequently, the following analysis or classification based on the connectivity matrix (i.e., comprising connectivity strengths of all pairs of brain regions or channels) can be conducted to reveal brain connectivity patterns or to detect cognitive states or brain diseases. The methods frequently used for estimating connectivity strengths are detailed in the following subsections except the last subsection where graph theoretical metrics are descibed.  

\subsection{Phase Locking Value}     
Phase locking value (PLV) shows how two signals measured from two separate brain regions synchronize in the phase. Let $s_k(t)$ and $s_l(t)$ indicate signals measured from brain regions $k$ and $l$, respectively. The analytical representations of $s_k(t)$ and $s_l(t)$ are obtained by the Hilbert transform, expressing as \cite{tass1998detection,  celka2007statistical, aydore2013note}
\begin{equation}
\begin{array}{l}
{z_k}(t) = {A_k}(t){e^{j{\phi _k}(t)}}\\
{z_l}(t)\; = {A_l}(t){e^{j{\phi _l}(t)}}.
\end{array}
\end{equation}
Phase differences at each time point are then calculated by
\begin{equation}
\Delta {\phi_{k,l}} (t) = {\phi _k}(t) - {\phi _l}(t).
\end{equation}
Subsequently, the PLV between the brain region $k$ and the brain region $l$ is obtained by averaging over all time points
\begin{equation}
PLV(k,l) = \frac{1}{n_t} \mid {\sum\limits_{t = 1}^{n_t} {{e^{j\Delta {\phi _{k,l}}(t)}}} } \mid,
\end{equation}
where $n_t$ is the number of time points. PLV is within the range $[0, 1]$ with 1 reflecting perfect phase synchronization between brain regions and 0 reflecting no phase synchronization. The PLV calculation is repeated for all pairs of brain regions. After that, all PLVs are assembled to form a connectivity matrix.  

\subsection{Phase Lag Index} 
When PLV is applied to EEG signal, spurious correlations between EEG signals are introduced into connectivity estimation due to that EEG signals from nearby electrodes are very likely to record the same brain activity source (a.k.a., common source). This is referred to as volume conduction. If two signals measure the same source, their phases should be well coupled. Therefore, a consistent and non-zero phase lag between signals cannot be attributed to that the signals are from the same source. The method measuring consistent and non-zero phase lag is called phase lag index (PLI) \cite{Stam2007}. PLI can be calculated based on phase differences at each time point  $\Delta {\phi _{k,l}}({t}),\; t\in\{ t_1, t_2, \cdots t_n\}$ by
\begin{equation}
PLI(k,l) = \left| {\left\langle {sign[\Delta {\phi _{k,l}}({t})]} \right\rangle } \right|, -\pi<\Delta {\phi}\leq\pi
\end{equation}
where $\left\langle \boldsymbol{\cdot} \right\rangle $ denotes the mean averaged over time $t\in\{ t_1, t_2, \cdots t_n\}$, $\left| \boldsymbol{\cdot} \right|$ denotes the absolute value, and  $sign$ stands for signum function. Similar to the PLV, PLI value ranges from 0 to 1. A value of 0 reflects either no synchronization or phase synchronization difference centered around 0 and $\pi $ while a value of 1 reflects perfect phase synchronization with consistent phase difference other than 0 and $\pi $ \cite{Stam2007}. 

\subsection{Wavelet Phase Coherence}
Let ${\phi _x}(t,f)$ and ${\phi _y}(t,f)$ be phases at time point $t$ and frequency $f$ for signals $x(t)$ and $y(t)$, respectively. The phase difference can then be calculated by
\begin{equation}
\Delta \phi (t,f) = {\phi _x}(t,f) - {\phi _y}(t,f)
\end{equation}
Wavelet phase coherence is defined in terms of $\cos \;\Delta \phi (t,f)$ and $\sin \;\Delta \phi (t,f)$
\begin{equation}
V(f) = \sqrt {{{\left\langle {\cos \;\Delta \phi (t,f)} \right\rangle }^2} + {{\left\langle {\sin \;\Delta \phi (t,f)} \right\rangle }^2}} 
\end{equation}
where $\left\langle \boldsymbol{\cdot} \right\rangle $ denotes the mean averaged over time \cite{bandrivskyy2004wavelet}. 

\subsection{Pearson Correlation and Cross Correlation}
Pearson correlation is a quantity measuring how a signal $x(t)$ correlates with another signal $y(t)$, defined as 
\begin{equation}
{\rho _{x,y}} = \frac{{E[(X - {\mu _x})(Y - {\mu _y})]}}{{{\sigma _x}{\sigma _y}}}
\end{equation}
where $\mu_x$ and $\mu_y$ are means of signals $x(t)$ and $y(t)$, respectively, and $\sigma_x$ and $\sigma_y$ are their corresponding standard deviations. \\

Cross correlation is defined as \cite{bracewell1986fourier}
\begin{equation}
\rho (\tau ) = \int {\bar x(t)} y(t + \tau )dt
\end{equation}
where $\bar x(t)$ denotes the complex conjugate of $x(t)$ and $\tau$ is the lag.

In the studies with fMRI, Pearson correlation is frequently adopted to estimate functional connectivity between voxels or between atlas-based regions or from a seed region to all other regions.  

\subsection{Mutual Information}
Mutual information between two signals is defined as \cite{cover2012elements}
\begin{equation}
{I_{x,y}} = H(x) + H(y) - H(x,y)
\end{equation}
where $H(x)$ and $H(y)$ are Shannon entropies of signals $x(t)$ and $y(t)$, respectively, and $H(x,y)$ is joint entropy. 

\subsection{Granger Causality} 
The aforementioned methods estimate connectivity strength for a pair of brain regions at a time. Besides, a multivariate autoregressive (MAR) model can be employed to simultaneously estimate relationships among more than two brain regions. Suppose that there are $N$ signals, represented by $X = {[\begin{array}{*{20}{c}}
{{x_1}}&{{x_2}}& \cdots &{{x_N}}
\end{array}]^{\rm T}}$. The current values at each signal are dependent on the past values at each signal, modelling as
\begin{equation}
\label{AR}
\left[ {\begin{array}{*{20}{c}}
{{x_1}(t)}\\
{{x_2}(t)}\\
 \vdots \\
{{x_N}(t)}
\end{array}} \right] = \sum\limits_{r = 1}^p {{A_r}} \left[ {\begin{array}{*{20}{c}}
{{x_1}(t - r)}\\
{{x_2}(t - r)}\\
 \vdots \\
{{x_N}(t - r)}
\end{array}} \right] + \left[ {\begin{array}{*{20}{c}}
{{w_1}(t)}\\
{{w_2}(t)}\\
 \vdots \\
{{w_N}(t)}
\end{array}} \right]
\end{equation}
where $W = {[\begin{array}{*{20}{c}}
{{w_1}}&{{w_2}}& \cdots &{w{}_N}
\end{array}]^{\rm T}}$ is white uncorrelated noise and $p$ is order of the MAR model. The model order $p$ can be determined by model quality assessment such as Akaike information criterion (AIC) \cite{akaike1974new}. The AIC makes a good balance between the goodness of fit of the model and the model simplicity.
${A_r},\;r \in \{ 1,\;2,\; \cdots ,\;p\}$ are $N \times N$ coefficient matrices for each lag $r$.
 
\begin{equation}
{A_r} = \left[ {\begin{array}{*{20}{c}}
{{a_{11}}(r)}&{{a_{12}}(r)}& \cdots & \cdots &{{a_{1N}}(r)}\\
{{a_{21}}(r)}& \vdots & \vdots & \vdots &{{a_{2N}}(r)}\\
 \vdots & \vdots & \vdots &{{a_{ij}}(r)}& \vdots \\
 \vdots & \vdots & \vdots & \vdots & \vdots \\
{{a_{N1}}(r)}&{{a_{N2}}(r)}& \cdots & \cdots &{{a_{NN}}(r)}
\end{array}} \right]
\end{equation}

The element ${{a_{ij}}(r)}$ in the coefficient matrix $A_r$ describes linear prediction effect of the $r$th past value ${x_j}(t - r)$ of the signal ${x_j}(t)$ on predicting ${x_i}(t)$. if any ${a_{ij}}(r) \ne 0$, it means that ${x_j}(t)$ Granger-causes ${x_i}(t)$ \cite{granger1969investigating}. It is worth noting that ${x_j}(t)$ Granger-causing ${x_i}(t)$ does not mean that ${x_i}(t)$ must Granger-cause ${x_j}(t)$. It is not reciprocal. Based on Granger causality, a few variants have been developed, which are introduced below.     

\subsection{Ordinary Coherence}
MAR model formulated in (\ref{AR}) can be rewritten as
 \begin{equation}
 \label{w}
W(t) = \sum\limits_{r = 0}^p {{{\bar A}_r}X(t - r)} 
\end{equation}
where 
\begin{equation}
{\bar A_r} = \left\{ {\begin{array}{*{20}{c}}
{I,\;\;\;\;r = 0}\\
{ - {A_r},\;\;0 < r \le p}
\end{array}} \right.
\end{equation}
Equation (\ref{w}) can be transformed into frequency domain as
  \begin{equation}
X(f) = {\bar A^{ - 1}}(f)W(f)
\end{equation}
where
\begin{equation}
\bar A(f) = \sum\limits_{r = 0}^p {{{\bar A}_r}{z^{ - r}},\;z = {e^{ - 2\pi if}}} 
\end{equation}
The cross-spectral power density matrix obtained by
\begin{equation}
S(f) = E[X(f)X{(f)^H}] = H(f)\Sigma H{(f)^H}
\end{equation}
where the superscript $H$ represents Hermitian transpose, 
\begin{equation}
\Sigma = \left[ {\begin{array}{*{20}{c}}
{\sigma _{11}^2}&{{\sigma _{12}}}& \cdots &{{\sigma _{1N}}}\\
{{\sigma _{21}}}&{\sigma _{22}^2}& \cdots &{{\sigma _{2N}}}\\
 \vdots & \vdots & \ddots & \vdots \\
{{\sigma _{N1}}}& \cdots & \cdots &{\sigma _{NN}^2}
\end{array}} \right]
\end{equation}

stands for covariance matrix, and
  \begin{equation}
H(f) = {\bar A^{ - 1}}(f)
\end{equation}

Ordinary coherence is defined as \cite{bendat2011random}
\begin{equation}
{C_{ij}}(f) = \frac{{{{\left| {{S_{ij}}(f)} \right|}^2}}}{{{S_{ii}}(f){S_{jj}}(f)}}
\end{equation}
It shows the extent to which brain regions $i$ and $j$ are simultaneously activated. 

\subsection{Directed Coherence}
Directed coherence, unlike ordinary coherence merely describing mutual synchronicity, gives both connective strength and connective direction between brain regions. The directed coherence from brain region $j$ to brain region $i$ can be expressed as \cite{schnider1989detection, baccala1998studying, kaminski1991new}
\begin{equation}
{\gamma _{ij}}(f) = \frac{{{\sigma _{jj}}{H_{ij}}(f)}}{{\sqrt {\sum\nolimits_{j = 1}^N {\sigma _{jj}^2{{\left| {{H_{ij}}(f)} \right|}^2}} } }}
\end{equation}

\subsection{Partial Directed Coherence} 
Partial directed coherence (PDC) from brain region $j$ to brain region $i$ is defined as \cite{kaminski12005causal, baccala2001partial}
\begin{equation}
{\pi _{ij}}(f) = \frac{{\bar A{}_{ij}(f)}}{{\sqrt {\sum\nolimits_{i = 1}^N {{{\left| {\bar A{}_{ij}(f)} \right|}^2}} } }}
\end{equation}
 
\subsection{Directed Transfer Function}
Directed transfer function is defined in terms of $H$ as
\begin{equation}
\gamma _{ij}^2(f) = \frac{{{{\left| {{H_{ij}}(f)} \right|}^2}}}{{\sum\nolimits_{j = 1}^N {{{\left| {{H_{ij}}(f)} \right|}^2}} }}
\end{equation}
$\gamma _{ij}^2(f)$ quantifies the fraction of inflow to brain region $i$ stemming from brain regions $j$ \cite{kaminski1991new}.

\subsection{Full Frequency Directed Transfer Function}
After the frequency normalization, directed transfer function is transformed into full frequency directed transfer function, expressing as \cite{kaminski12005causal}
\begin{equation}
F_{ij}^2(f) = \frac{{{{\left| {{H_{ij}}(f)} \right|}^2}}}{{\sum\nolimits_f {\sum\nolimits_{j = 1}^N {{{\left| {{H_{ij}}(f)} \right|}^2}} } }}
\end{equation}

\subsection{Synchronization Likelihood}
Synchronization likelihood quantifies the extent to which a signal recorded from a brain region synchronized to signals from all the other brain regions. 
Synchronization likelihood of signal $x(t)$ is defined as \cite{stam2002synchronization}
\begin{equation}
{S_k} = \frac{1}{{2(w2 - w1)}}\sum\limits_{i = 1}^t {\sum\limits_{\scriptstyle\;\;\;\;\;\;\;j = 1\hfill\atop
\scriptstyle w1 < \left| {j - i} \right| < w2\hfill}^t {{S_{k,i,j}}} } 
\end{equation}
where
\begin{equation}
{S_{k,i,j}} = \left\{ {\begin{array}{*{20}{c}}
{\frac{{\sum\limits_{k = 1}^N {\theta ({\varepsilon _{k,i}} - d({x_{k,i}} - {x_{k,j}}))}  - 1}}{{N - 1}},\;d({x_{k,i}} - {x_{k,j}}) < {\varepsilon _{k,i}}}\\
{0,\;\;\;\;\;\;d({x_{k,i}} - {x_{k,j}}) \ge {\varepsilon _{k,i}}\;\;\;\;\;\;\;}
\end{array}} \right.
\end{equation}
$w1$ and $w2$ are parameters to determine the time window where data points are included. $N$ is the number of signals, $\theta()$ is the Heaviside step function, $d()$ is the Euclidean distance, and $x_{k,i}$ is embedded vector reconstructed with time-delay embedding \cite{takens1981detecting}. 

\subsection{Graph Theoretical Metrics}
Once a connectivity matrix is constructed by any one of the aforementioned methods, connectivity attributes can be quantified by individual connectivity strengths representing relationships between any two brain regions or by graph theoretical metrics characterizing topological properties of brain connectivity network with all connections. In general, these metrics can be divided into two categories: local metric and global metric. Local metric quantifies topological organization within a brain area or part of a connectivity network. It includes nodal degree, nodal strength and betweenness centrality \cite{freeman1979centrality}, and so forth. Global metric quantifies connectivity network topology from the view of the whole network. The wildly-used global metrics are small-worldness \cite{humphries2008network}, global efficiency \cite{latora2001efficient}. For the metrics of clustering coefficient and characteristic path length \cite{watts1998collective, rubinov2010complex}, they can be either local metrics or global metrics depending on either computing on individual nodes or averaging over all nodes of a network. A brain network usually exhibits segregation and integration. The segregation can be quantified by network clustering coefficient, reflecting specialized processing in each brain region. The integration can be quantified by global efficiency, reflecting the whole communication/cooperation efficiency among distributed brain regions. If a network is of both densely interconnected localization and high entire efficiency, it is called small-world network.  The small-world network is considered as an optimal network with the balance of integration and segregation. For the details of graph theoretical metrics, readers may refer to the reference \cite{rubinov2010complex}. Graph theoretical metrics can be adapted to their dynamic version by taking temporal information into consideration when computing the metrics.

\section{Mental Fatigue}
\label{MentalFatigue}
Fatigue is divided into two categories: physical fatigue and mental fatigue. We only discuss mental fatigue in this chapter. Mental fatigue is usually caused by excessive demand on brain resource for implementation of mental tasks or prolonged task execution with exhaustive depletion of brain resource or prolonged monotonous task reducing attention retention. As we know, mental fatigue would lead to a decline in brain function, showing inefficient performance, increased error occurrence rate, and inability to continue the ongoing task, and so forth. It, sometimes, even results in fatal consequences such as deadly traffic accident \cite{connor2002driver}. Therefore, understanding the neural mechanisms under mental fatigue and the characteristics during the transition from wakefulness to fatigue are crucial to inform the development of precautionary measures and prevent adverse consequences stemming from mental fatigue. Up to now, numerous studies have been conducted to reveal mental fatigue-related effects on the human brain. We only collected and listed studies that investigated or classified mental fatigue from the perspective of brain connectivity in Table \ref{table1} as this chapter focuses on the theme of brain connectivity. The majority of studies analysed either EEG data or fMRI data to explore neural mechanisms in brain connectivity pertaining to mental fatigue \cite{Harvy2019, Boissoneault2016a, Zhang2017, Hampson2015, Li2016, Sun2017, zhao2017reorganization,   Finke2015, Nordin2016, Kong2015}. Few others devoted to classifying mental fatigue state using features extracted from EEG signal \cite{Cynthia2017, Harvy2018}, as well as a combination of data analysis and classification \cite{Dimitriadis2013, wang2020driving}. The experimental settings, findings derived from studies and classification performances were detailed in the following subsections.                       

\begin{landscape}
\renewcommand{\arraystretch}{2}
\begin{center}
\begin{longtable}{p{1.5cm} p{3cm} p{1cm} p{3cm} p{4.5cm} p{1.5cm}}

\caption{Mental Fatigue Studies Exploring from the Perspective of Brain Connectivity}\label{table1}\\
\hline
\bfseries Reference&	\bfseries Category	& \bfseries Signal Modality&	\bfseries Method	& \bfseries Subject&	\bfseries Type\\
\hline
\endfirsthead

\multicolumn{6}{c}%
{\bfseries \tablename\ \thetable{} -- continued from previous page}\\
\hline \bfseries Reference &	\bfseries Category	& \bfseries Signal Modality &	\bfseries Method	& \bfseries Subject &	\bfseries Type\\ \hline
\endhead

\hline \multicolumn{6}{c}{Continued on next page} \\ \hline
\endfoot

\hline \hline
\endlastfoot

Boissoneault  et al. [a] \cite{Boissoneault2016a}&	Fatigue in patients with myalgic encephalomyelitis/chronic fatigue syndrome (ME/CFS)&	fMRI &	Seed-based functional connectivity&	17 patients with ME/CFS vs. 17 healthy controls&	Analysis\\
Boissoneault  et al. [b] \cite{Boissoneault2016}&	Fatigue in patients with ME/CFS&	fMRI&	Independent Component Analysis (ICA) and Seed-based functional connectivity&	19 patients with ME/CFS vs. 17 healthy controls&	Analysis\\
CruzGomez  et al. \cite{CruzGomez2013}&	Fatigue in patients with multiple sclerosis&	fMRI&	ICA and Seed-based functional connectivity&	32 fatigued patients vs. 28 non-fatigued patients vs. 18 healthy controls&	Analysis\\
Cynthia et al. \cite{Cynthia2017} &	Driving fatigue	& EEG	& Phase Locking Value (PLV) &	20 young subjects	& Classification \\
Chen et al. \cite{Chen2019}&   Real driving experiment&    EEG  & Phase lag index&  14  young healthy male subjects&   Analysis and Classification\\   
Dang et al. \cite{dang2020rhythm} &Driving simulation &EEG&Correlation&16 healthy subjects&Analysis and Classification\\
Dimitrakopoulos  et al. \cite{Dimitrakopoulos2017a}&	Driving fatigue&	EEG&	PDC, DTF, and PLI&	20 healthy subjects&	Classification\\
Finke  et al. \cite{Finke2015}&	Fatigue in patients with multiple sclerosis&	fMRI and DTI &	Seed-based functional connectivity&	44 multiple sclerosis patients with fatigue vs. 20 healthy controls&	Analysis\\
Hampson  et al. \cite{Hampson2015}&	Fatigue in patients with persistent breast cancer&	fMRI &	ICA and Seed-based functional connectivity&	15 breast cancer patients with vs. 8 breast cancer patients without fatigue&	Analysis\\
Harvy et al. \cite{Harvy2018}&	Driving fatigue&	EEG&	Partial directed coherence (PDC)&	30 healthy subjects&	Classification\\
Harvy et al. \cite{Harvy2019} & Driving fatigue & EEG &Pearson correlation & 30 healthy subjects & Analysis\\
Kong et al. \cite{Kong2015}&	Driving simulation&	EEG&	Granger causality&	12 healthy subjects &	Analysis\\
Kar et al. \cite{Kar2011}&	Physical exercise, simulated driving, driving-related computerised game (only driving-related computerised game was reported)&	EEG&	synchronisation likelihood&	12 healthy subjects&	Analysis\\
Liu et al. \cite{Liu2010a}&	Mental fatigue induced by cognitive task&	EEG&	Directed transfer function (DTF)&	50 healthy male subjects&	Analysis\\
Nordin  et al. \cite{Nordin2016}&	Psychomotor vigilance task (PVT), mild traumatic brain injury with persisting symptons of fatigue&	fMRI&	Voxel-wise cross-correlation coefficient &	10 patients with mild traumatic brain injury fatigue vs. 10 healthy controls&	Analysis\\
Qi et al. \cite{Qi2020} &   Visual oddball task&    fMRI&   Pearson correlation&      20 healthy subjects&	Analysis\\
Ramage et al. \cite{Ramage2019}&   Constant effort task&    fMRI&   Pearson correlation&     60 patients with mild traumatic brain injury vs.  42 controls &Analysis\\
Sun et al. [a] \cite{Sun2014a}&	PVT&	EEG&	 PDC&	26 young subjects&	Classification\\
Sun et al. [b] \cite{Sun2014}&	PVT&	EEG&	PDC (8-10Hz)&	32 subjects (6 subjects excluded due to poor motivation on the task)&	Analysis\\
Sun et al. [c] \cite{Sun2017}&	Visual oddball task&	fMRI&	Pearson correlation&	20 healthy subjects&	Analysis\\
Wang et al. \cite{Wang2015a}&	Realistic Driving&	EEG&	Fuzzy synchronization likelihood &	20 healthy subjects (4 subjects excluded due to poor data acquisition)&	Analysis and Classification\\
Wang et al. \cite{wang2020driving}&   Driving simulation& EEG&    Phase lag index&   20 healthy subjects&      Analysis and Classification\\				
Xu et al. \cite{Xu2017}&	Driving with and without mental calculation&	fNIRS &	wavelet coherence and wavelet phase coherence&	14 healthy young subjects&	Analysis\\
Zhang et al. \cite{Zhang2017}&	Fatigue in patients with Parkinson's disease (PD)&	fMRI&	Seed-based functional connectivity&	17 PD patients with fatigue vs. 32 PD patients without fatigue vs. 25 healthy controls&	Analysis\\
Zhang et al. \cite{zhang2020spatio} &Driving simulation &EEG& Correlation&12 subjects&Analysis\\
Zhao et al. \cite{zhao2017reorganization} &	Driving fatigue and visual oddball task	&EEG	&Ordinary Coherence (OC)		&	16 healthy young subjects &	Analysis\\

\end{longtable}
\end{center}
\end{landscape}

\subsection{Tasks Conditions in Fatigue Studies}
\begin{figure}[htb]
 \centering
 \includegraphics[width=0.9\textwidth]{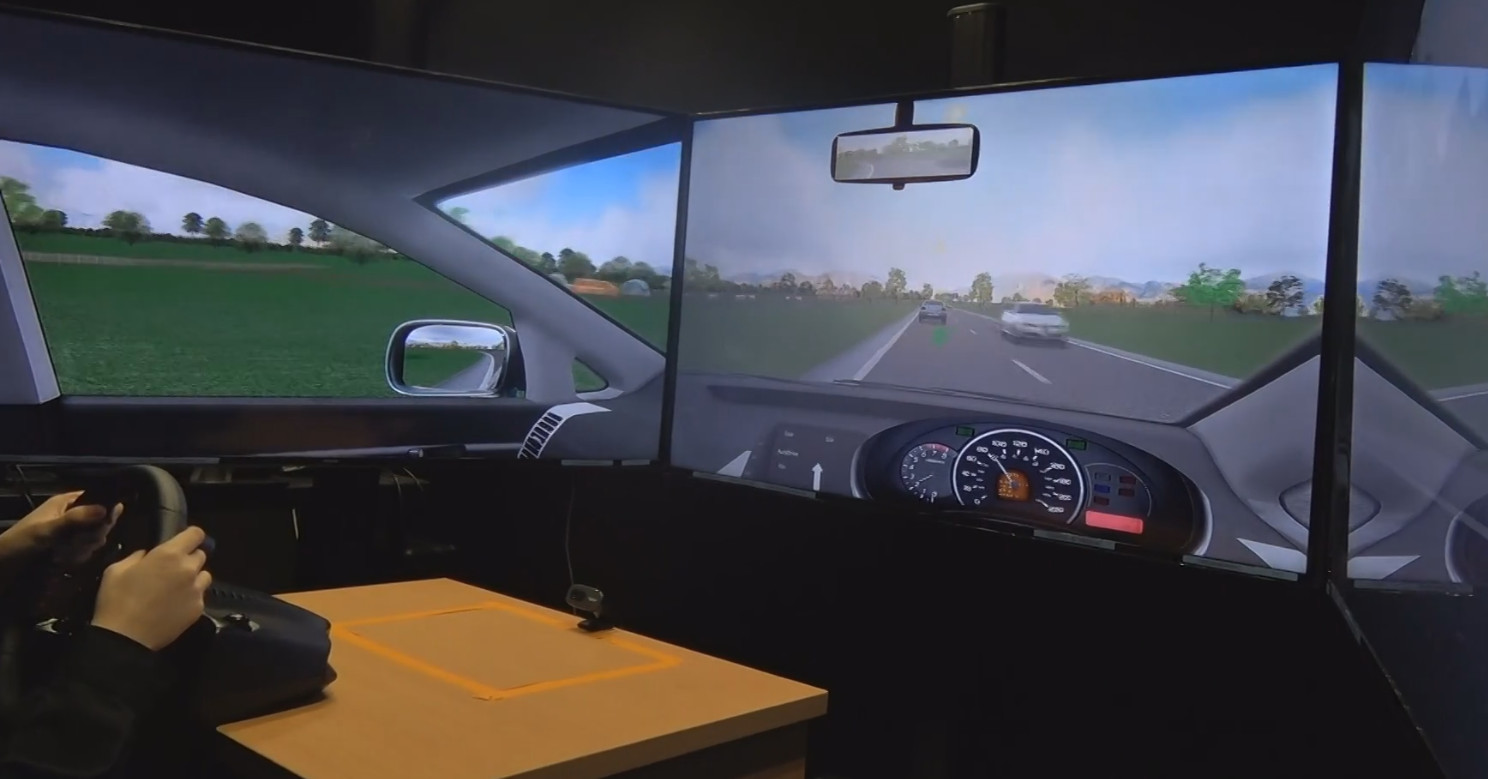}
 \caption{Illustration of driving simulation at an experiment for mental fatigue investigation.}
 \label{DS}
 \end{figure}

Most frequently, simulated driving is employed to induce fatigue because the task can be well defined and be easily controlled in a simulated environment (see Fig. \ref{DS}). It also has an advantage of safety and participants do not face any risks of physical injury. Responses and task difficulty can be adjusted to meet the requirements of a study. In such a driving simulation experiment, participants usually undergo a certain long time of simulated driving to attain fatigue state \cite{bose2019regression, Cynthia2017, Harvy2018, Kong2015, wang2020driving}. Inducing fatigue is much quicker and easier when highly demanded attention is involved in tasks such as psychomotor vigilance task (PVT) \cite{Nordin2016, Sun2014a}. During the PVT, participants should respond to a counter as quickly as they can and intense attention is required to perform the task. This imposes high mental load on the human brain and swiftly depletes brain resource, which facilitates to induce mental fatigue. Another attention-required task that has been used for fatigue study is visual oddball task \cite{Li2016, Sun2017}, in which participants are asked to distinguish the target letter from shape-analogous English letters by pressing a predefined bottom. Moreover, other attention-based tasks such as detection of three different odd numbers and responses according to image color and arithmetic calculation such as addition and subtraction have also been utilized for fatigue studies \cite{Liu2010a}. Besides these laboratory-based experiments, real outdoor driving lasting about 3.5 hours was conducted to explore mental fatigue \cite{Wang2015a}. In addition, fatigue is  investigated not only in healthy people but also in patients. Patients with fatigue symptoms are compared to those who have no such fatigue symptoms to reveal fatigue effect on diseases (e.g., Parkinson's disease and multiple sclerosis) \cite{Boissoneault2016a, Boissoneault2016, Zhang2017, Finke2015, Hampson2015, Nordin2016, Ramage2019}.

\subsection{Findings Derived from Analysis}
\begin{figure}[tbh]
 \centering
 \includegraphics[width=0.9\textwidth]{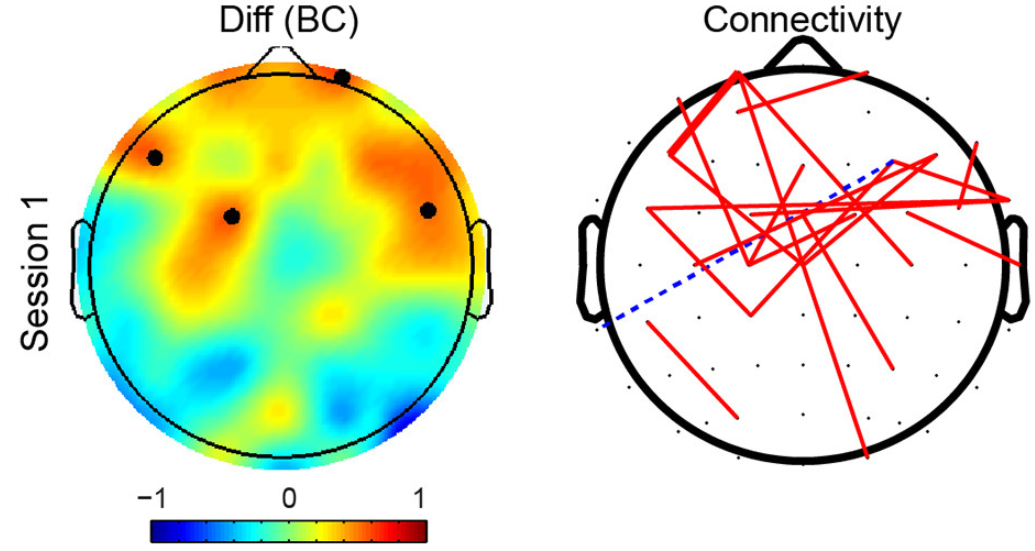}
 \caption{Betweenness centrality difference between after and before the consecutive visual oddball task. Black dots stand for EEG channels with significant changes and lines represent dominant connections. Red color means the increase after the task while blue color means the decrease after the task. Brain regions residing in the frontal area appeared drastic changes in the connectivity after the task. This figure was reproduced from \cite{Li2016} with the reuse permission under an open licence.}
 \label{fatigue}
 \end{figure}

Generally speaking, mental fatigue is related to overall low functional connectivity among brain regions. Denser local connectivity (within local communities) and lower long-distance connectivity (between communities) are observed in the state of mental fatigue. These contribute to inefficient connectivity for the whole brain network, resulting in low global efficiency. With relatively dense 64-channel EEG recording, increased clustering coefficient, increased characteristic path length, increased local efficiency, and decreased global efficiency in topological organization of connectivity network after consecutive implementation of a visual oddball task were found \cite{Li2016}, suggesting task implementation towards mental fatigue led to increased activity within local areas and fewer interactions between areas. The regional analysis using betweenness centrality revealed that brain regions located in the frontal area appeared drastic changes in the connectivity (see Fig. \ref{fatigue}). The similar changes in clustering coefficient and characteristic path length were also observed in a study with a simulated driving task \cite{zhao2017reorganization}. These results demonstrated that different tasks could lead to a similar topological change in brain connectivity network. It is worth mentioning that this is not always the case. In a study using sleep deprivation for inducing fatigue and synchronisation likelihood for estimating brain inter-regional connectivity, decreased characteristic path length was observed when a longer time for sleep deprivation was administered (more fatigued after the longer time for sleep deprivation) \cite{Kar2011}. This discrepancy might be attributed to different methods used for estimating brain inter-regional connectivity and a small number of electrodes (i.e., 19 electrodes) employed for EEG data collection in the study. With a relatively large cohort of 50 participants, Liu et al. found that the coupling among frontal, parietal, and central areas was altered from pre- to post-mental task, and has asymmetric changes in the left and right hemispheres \cite{Liu2010a}. The hemispheric asymmetry of functional connectivity was also observed in the frontoparietal area according to the study with PVT task \cite{Sun2014}. Using data modality of functional near-infrared spectroscopy (fNIRS), significantly lower wavelet coherence/wavelet phase coherence was observed in the prefrontal cortex in the frequency bands of 0.6$\sim$2 Hz and 0.052$\sim$0.145 Hz and the motor cortex in the frequency band of 0.021$\sim$0.052 Hz after the driving task \cite{Xu2017}. The involvement of the frontal area in the mental fatigue was also reported by studies using other modalities such as EEG \cite{Kong2015} and fMRI \cite{Sun2017}. Taken together, we can see that the frontal area plays an important role in brain connectivity alterations relevant to mental fatigue.  

In addition to studies in healthy people, fatigue was usually investigated with diseases. Patients with myalgic encephalomyelitis/chronic fatigue syndrome (ME/CFS) exhibited altered functional connectivity of several regions (e.g., bilateral superior frontal gyrus and precuneus) \cite{Boissoneault2016a}. The altered extent was correlated with clinical fatigue ratings. This was corroborated by another study from the same group \cite{Boissoneault2016}. In the comparison between the group of Parkinson's disease with fatigue and the group of Parkinson's disease without fatigue, it revealed that Parkinson's disease-related fatigue was associated with altered functional connectivity among the right middle frontal gyrus, left insula, and right midcingulate cortex \cite{Zhang2017}. Finke et al. investigated fatigue in the disease of multiple sclerosis and found that functional connectivity of basal ganglia nuclei with medial prefrontal cortex, precuneus and posterior cingulate cortex was negatively correlated with fatigue severity while the functional connectivity between caudate nucleus and motor cortex was positively correlated with fatigue severity \cite{Finke2015}. The association between fatigue in multiple sclerosis and functional connectivity was also discovered from another independent study \cite{CruzGomez2013}. According to the research on breast cancer, fatigued patients (i.e., persistent cancer-related fatigue) displayed greater connectivity from the left inferior parietal lobule (IPL) to superior frontal gyrus and lower connectivity from the right precuneus to the periaqueductal gray and from the left IPL to subgenual cortex \cite{Hampson2015}. As similar use in healthy people, the PVT was utilized to explore fatigue in patients with mild traumatic brain injury. A significant linear correlation between self-reported fatigue and functional connectivity in the thalamus and middle frontal cortex was found \cite{Nordin2016}. These studies demonstrated that fatigue-related changes in functional connectivity exist with diseases.          
            
\subsection{Fatigue Classification}
One step closer to practical usage is fatigue classification based on functional connectivity. Fatigue levels can be classified and an appropriate warning is given to drivers or operators whose fatigue extent reaches a certain high level. Most of fatigue classification papers aimed to detect driving fatigue in order to eliminate or reduce fatigue-caused traffic accidents \cite{Harvy2018, Cynthia2017, Dimitrakopoulos2017a, he2018boosting, wang2018novel, Wang2015a}. Others are for the purposes of separation of wakefulness and fatigue states during a task \cite{Sun2014a}. Connective strengths estimated by the methods presented in Sect. \ref{1.2} (e.g., PLI and PDC \cite{Dimitrakopoulos2017a}) are usually used as feature candidates. Feature selection follows to screen feature candidates, retaining discriminative features while excluding those features without value in classification. Feature selection can be done by evaluating individual features separately \cite{li2016decoding} or considering a subset of features at a time \cite{Dimitrakopoulos2017a} or using dimension reduction algorithms \cite{Harvy2018}. After that, a classifier named support vector machine is frequently employed to classify fatigue levels based on the selected features \cite{Dimitrakopoulos2017a, Sun2014a, Cynthia2017}. Besides, classifiers, such as linear discriminant analysis, k-nearest neighbours, sequential minimal optimization, least squares learning, and artificial neural network, have been utilized for fatigue classification \cite{Dimitrakopoulos2017a, Cynthia2017}. Recently, the fusion of connectivity features and power spectral density features was proposed to classify driving fatigue \cite{Harvy2018}. The classification performance was improved as complementary information was taken in the feature fusion. Mostly, k-fold cross-validation is used to obtain classification accuracy. The procedure of the k-fold cross-validation is as follows:
\begin{enumerate}
	\item Shuffle samples randomly
	\item Split the samples into $k$ groups (e.g., $k=10$)
	\item Take a group as the testing dataset
	\item Take the remaining $k-1$ groups as the training dataset
	\item Train a model using the training dataset and evaluate the model on the testing dataset
	\item Retain the accuracy obtained from the evaluation on the testing dataset and take another group as the testing dataset to repeat above steps 4-5 until each group has been the testing dataset once.
	\item Average accuracies across $k$ groups to obtain final classification accuracy
\end{enumerate}                 
Each sample is only used in the testing dataset once and used to train the model $k-1$ times. The more folds split, the more inflated classification performance. In an extreme case, one sample forms a group. It is called leave-one-out cross-validation (LOOCV). The LOOCV results in an inflated performance, which is an optimistic estimate of classification accuracy. To circumvent this issue, samples should be split into a few folds such as 5-fold and 10-fold \cite{li2018machine}. Noting that the number of folds depends on sample size. If samples are more, they could be split into a more number of folds. Due to the truth that classification performance is largely dependent on dataset itself and evaluation setting, classification accuracies reported in different studies cannot be directly compared. Therefore, we do not summarize classification accuracies and compare them in this chapter.          
 
\section{Mental Workload}
\label{MentalWorkload}
\begin{figure}[htb]
 \centering
 \includegraphics[width=0.9\textwidth]{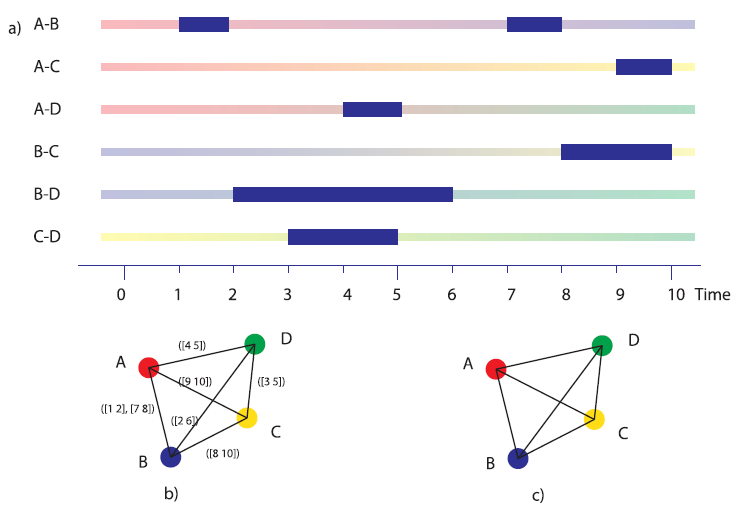}
 \caption{Illustration of the dynamic network. (a) A four-node (A, B, C, and D) network with the highlighted periods (in dark blue) of connectivity between any two nodes. (b) Dynamic graph representation of the network with the time information of connections shown on the edges. (c) The static graph representation of the network. Reprinted from \cite{Ren2017} with permission of IEEE.}
 \label{dynamicnetwork}
 \end{figure}

Mental workload is an amount of brain resource paid for task implementation. There has been no clear and universally-accepted definition for the mental workload. It is closely related to two determinants: task requirement and human capability or resources \cite{welford1978mental, wilson1991psychophysiological, huey1993workload, gopher2013analysis, gopher1986workload}. Given these two determinants, mental workload could refer to the portion of information processing capacity or resources that is actually required to meet task demands \cite{wilson1991psychophysiological} or mental workload could be viewed as the ratio of the resources demanded for satisfying task performance to the mental capacity a person can supply \cite{gopher1986workload}. Unlike mental fatigue that cannot be changed in a transient period, the mental workload could be modulated instantly. N-back (it requires participants to indicate whether the current stimulus matches the one from N steps earlier in a sequence of stimuli.) is one of the commonly used tasks to induce different levels of workload by giving different difficulty levels. A study comparing the 0-back and 2-back revealed that a higher functional integration in the theta band and lower functional segregation in the alpha band for the 2-back compared to the 0-back. The regions with significant differences between the 0-back and 2-back mainly resided in the frontal, temporal, and occipital lobes \cite{Dai2017}. Functional connectivity research on mental workload has been extended to the dynamic network, which takes temporal information into consideration when estimating connectivity strengths between brain regions (see Fig. \ref{dynamicnetwork} for the illustration). Ren et al. found that the higher workload led to a more globally efficient but less clustered dynamic small-world functional network based on dynamic network \cite{Ren2017}. 

Similar to the N-back's principle of that difficulty level is elevated by increasing the information to be remembered, Charbonnier et al.'s study adjusted workload level through changing the length of sequential digits that participants have to remember for correctly making the following judgement \cite{Charbonnier2016}. They extracted connectivity features from the selected subset of electrodes and classified workload levels using Fisher's linear discriminant analysis (LDA). Moreover, arithmetic was utilized to induce different workload levels by modulating addition difficulty. Five workload levels induced by five arithmetical difficulties were classified by k-nearest neighbours (k-NN) classifier using different feature extraction methods (i.e., tensor subspace analysis (TSA) and LDA) and multiple bands on spatial dependent representations of phase synchronization. According to the report, a combination of TSA and k-NN using coupling features between the theta in the frontal region and the upper alpha (10$\sim$13 Hz) in parieto-occipital region achieved the best performance \cite{Dimitriadis2015}. In the study of using intelligence quotient test for workload induction, multiple features including connectivity metrics were fed into a neural network to predict workload level \cite{Friedman2019}. In addition, two different categories of mental tasks were used to induce mental workload. As addressed in \cite{Dimitrakopoulos2017}, N-back (0-back and 2-back) and arithmetic (one-digit number addition and three-digit number addition) tasks were used and both within-category classification and between-category classification were performed. A good performance was obtained for both classification cases. The relationship between functional connectivity and mental workload was observed in multiple independent studies \cite{Kosti2018, shaw2019cerebral, kakkos2019mental, pei2020eeg}. In a recent study, a regression model was used to establish the relationship between functional connectivity features and workload level in a programmer who performed code comprehension and syntax error detection \cite{Kosti2018}. This model can be used to evaluate the programmer's workload. The functional connectivity featured was also fused with power spectral density features to improve workload identification as shown in the study of multi-class workload identification based on a flight simulation experiment \cite{pei2020eeg}. Besides the tasks with only one participant, workload has been investigated in the case of cooperation between participants. Sciaraffa et al. explored workload among participants who engaged in a collaborative task and found that causality value was higher for those who collaborated more tightly \cite{Sciaraffa2019}.       
                   
\section{Vigilance}
\label{Vigilance}
Vigilance is defined as the ability to maintain concentrated attention and to remain alert to stimuli or targets over prolonged periods of time \cite{parasuraman1986vigilance, warm2008vigilance}. Vigilance is closely related to fatigue. Fatigue state is accompanied with low vigilance, but the non-fatigue state does not have to be with high vigilance. It has been found that vigilance was relevant to functional brain connectivity \cite{Teng2019}. An fMRI study revealed that vigilance was associated with the fluctuations in left amygdala functional connectivity with regions of the salience network \cite{Baczkowski2017}. Piantoni et al. used Granger causality to investigate cingulate functional connectivity and found that the forward effective connectivity over the cingulate was related to vigilance \cite{Piantoni2013}. When vigilance is degraded, functional connectivity between brain regions is disrupted  \cite{Abbasi2019}. The linkage between functional connectivity and vigilance was also supported by the finding derived from the analysis of dynamic functional connectivity \cite{Wang2016}. Because of the relevance between vigilance and functional connectivity, features derived from functional connectivity can be utilised to recognise vigilance. For example, Xie et al. suggested that the information derived from the alpha band networks could be used to predict vigilance level \cite{Xie2020}. It has been shown that phase synchrony indices on connectivity network can be used to successfully predict the average hit response time (a measure indicating vigilance) based on deep neural network model (this is a machine learning method) \cite{Torkamani-Azar2019}. An extended question might come up. Is it possible to enhance vigilance when it is declined? To this end, some efforts have been done. In the stroop color-word task, participants' vigilance was enhanced when an auditory stimulus was provided according to behavioural and connectivity analysis \cite{al2019brain}. The vigilance enhancement could also be attained by visual stimulus \cite{bodala2016eeg}.      
   
\section{Emotion}
\label{Emotion}
Emotion is an affective state of consciousness, which is associated with thoughts, feelings, behavioural responses, and interactions with others \cite{ekman1994nature}. Emotion can be characterized by two dimensions of arousal and valence. It is basically divided into six categories: anger, disgust, fear, happiness, sadness and surprise. Images of facial expressions or videos with emotional components (e.g., sounds) are used as stimuli in experiments for emotion studies. According to a study with a large number of participants (i.e., 586), connective strength was increased from amygdala to the hippocampus during the encoding of positive and negative pictures in relation to neutral pictures. The strength of this connection in the reverse direction showed a smaller elevation \cite{Fastenrath2014}. Diseases, such as major depressive disorder (MDD), affect the functional connectivity relevant to emotional stimuli. As indicated in the study of functional connectivity comparison between healthy group and MDD group, greater functional connectivity between the subgenual anterior cingulate cortex and the amygdala and lower functional connectivity between the subgenual anterior cingulate cortex and the insula/putamen, fusiform gyrus, precuneus/posterior cingulate, and middle frontal gyrus were found in the MDD group \cite{Ho2014}. A comparison study showed that functional connectivity was obviously different between positive emotion and negative emotion \cite{sorinas2020cortical}. This suggested that emotion can be recognised based on functional connectivity. As shown in the emotion recognition study, a classifier of quadratic discriminant analysis was used to successfully distinguish different emotional states based on connectivity indices such as between-channel coherence and correlation \cite{LeeYYClassifying}. The between-channel connective strengths were adopted to distinguish high arousal from low arousal, or positive valence from negative valence by traditional machine learning methods (e.g., support vector machine) \cite{Jahromy2019}. Emotion recognition was also achieved by more sophisticated methods such as graph convolutional neural network \cite{Wang2019}. According to recent studies, functional connectivity features were better than power spectral density features in emotion recognition \cite{Wu2019}. By combining these different kinds of features, emotion recognition performance can be improved compared to that of using single kind of features \cite{Al-Shargie2019a, Li2019}. It is not surprising because feature combination could provide more information that single kind of features cannot provide. In particular, performance improvement is predominant when different kinds of features are complementary.   

\section{Toolboxes for Brain Connectivity Analysis and Classification}
\label{Toolbox}
We introduce some toolboxes that can be used to facilitate your analysis and classification based on brain connectivity (see Table \ref{ToolboxesTable}). The toolboxes introduced for classification is universal, rather than only for brain connectivity-based classification. First of all, raw data need to be preprocessed to remove artifacts that are not interesting for the study. EEGLAB \cite{delorme2004eeglab} is one of the good choices you can choose to accomplish this purpose for time series data (e.g., EEG). It provides variant data import functions for you to import diverse data formats. The accompanying manual guides you step by step to process data. It can be easily used even you are a novice to the EEGLAB. For the fMRI and DTI images, the Statistical Parametric Mapping (SPM) package (http://www.fil.ion.ucl.ac.uk/spm/software), FMRIB Software Library (FSL) \cite{smith2004advances} can be utilized to process them, respectively. In addition to the self-contained functions in the above toolboxes for connectivity strength estimation, a specialized toolbox, called HERMES, is available at http://hermes.ctb.upm.es/ for strength estimation. It provides a user-friendly GUI to allow users to select methods (refer to Sect. \ref{1.2} for methods) for strength estimation. If you plan to explore dynamic brain connectivity, the toolbox named dynamic brain connectome (DynamicBC) can meet your need (available at http://www.restfmri.net/forum/DynamicBC). Once a connectivity matrix containing connective strength values for all pairs of channels/brain regions/electrodes is constructed, the brain connectivity toolbox (BCT) can be utilized to calculate graph theoretical metrics \cite{rubinov2010complex}. Please refer to the paper \cite{rubinov2010complex} for the definition and interpretation of the metrics. GRETNA is another toolbox to perform topological analysis of brain connectivity \cite{wang2015gretna}. To visualize results derived from brain connectivity analysis, BrainNet Viewer \cite{xia2013brainnet} and EEGNET \cite{hassan2015eegnet} are useful to display resultant connectivity pattern vividly. Besides, FieldTrip \cite{oostenveld2011fieldtrip} is a general-purpose toolbox for the processing, source localization, and analysis of neurophysiological signals (time series data).  

\begin{table}[h!]
    \caption{Toolboxes for Brain Connectivity Analysis and Classification}
    \label{ToolboxesTable}
    \begin{threeparttable}
    \begin{tabular}{lll} 
      \hline
      \textbf{Toolbox Name} & \textbf{Primary Usage} & \textbf{URL}\\
      \hline
      EEGLAB & EEG Processing and Analysis & https://sccn.ucsd.edu/eeglab/index.php\\
      SPM & fMRI Images Processing & http://www.fil.ion.ucl.ac.uk/spm/software\\
      FSL & DTI Images Processing & https://fsl.fmrib.ox.ac.uk/fsl/fslwiki/\\
      HERMES &Connectivity Strength Estimation&http://hermes.ctb.upm.es/ \\
      DynamicBC&Dynamic Connectivity Estimation&http://www.restfmri.net/forum/DynamicBC\\
      BCT&Computing Graph Metrics&https://sites.google.com/site/bctnet/\\
      GRETNA&Computing Graph Metrics&https://www.nitrc.org/projects/gretna/\\
      BrainNet Viewer&Brain Network Visualization&https://www.nitrc.org/projects/bnv/\\
      EEGNET&Brain Network Visualization&https://sites.google.com/site/eegnetworks/home\\
      FieldTrip&Neurophysiological Signal Processing&http://www.fieldtriptoolbox.org/\\
      LIBSVM&SVM Algorithm&https://www.csie.ntu.edu.tw/~cjlin/libsvm/\\
      Matlab SMLT&Basic Machine Learning Algorithms&https://www.mathworks.com/products/statistics.html\\
      Spider&Machine Learning Algorithms&http://people.kyb.tuebingen.mpg.de/spider/\\
      \hline\hline
    \end{tabular}
  \begin{tablenotes}
        \footnotesize
        \item[*] Only primary usage is listed in the table for each toolbox. Some toolboxes might have the functions that are not covered by the primary usage.
        \item[*] The URLs were tested in August 2020 and they all worked at the time of testing.
      \end{tablenotes}
    \end{threeparttable}
\end{table}

Support vector machine (SVM) is one of the excellent classifiers with good-balance between performance and complexity. A toolbox named LIBSVM was developed to easily implement SVM with parameter tuning \cite{fan2005working}. It can be downloaded at\\ 
https://www.csie.ntu.edu.tw/~cjlin/libsvm/. You may use Matlab built-in Statistics and Machine Learning Toolbox (SMLT) to perform classification. It contains basic methods for the classification. A comprehensive toolbox for classification including feature selection methods and discriminative methods is the Spider (available at http://people.kyb.tuebingen.mpg.de/spider/). It contains SVM, C4.5, k-NN and so on, as well as regression methods. You may select different methods and compare them in terms of classification accuracy.                                   

\section{Thoughts and Further Directions}  
\label{TFD}
Based on the progress of understanding of cognitive states and the development of classification technique for distinguishing cognitive states, it can be seen that neuroimaging data contain abundant information relevant to cognitive states and brain connectivity analysis using noninvasive recorded data is a feasible manner to reveal neural mechanisms underlying cognitive states. Cognitive state is investigated using experiments, which are designed to induce required mental state while excluding interferential factors as many as possible. However, absolute exclusion seems impossible and special caution should be given when interpreting results. Moreover, artifacts and non-cognitive state background give a confounding effect on findings. Therefore, preprocessing is a critical step before data analysis and data classification for removing the confounding effect. Until now, many preprocessing methods have been developed to remove diverse kinds of artifacts such as EOG originated from eye movements and EMG originated from head movements. The principle behind artifacts removal is that raw data are decomposed into components and components corresponding to artifacts are removed to eliminate artifacts. Independent component analysis (ICA) \cite{jung2000removing} principal component analysis (PCA) \cite{jung1998removing} and wavelet transform \cite{krishnaveni2006automatic} belong to this category. These methods are developed to process data which have been collected and are not intended to process data in real-time. Nonetheless, some of them have been modified to meet the requirement of real-time processing. For example, ICA method was modified to suit real-time data processing \cite{hsu2016real}. Albeit large progress has been made in the artifacts removal methodologies, new challenges of artifacts removal emerge as newly introduced recording techniques lead to new kinds of artifacts or artifacts mitigation is required in special situations. In recent years, concurrent EEG/fMRI recording \cite{he2008multimodal} is becoming more and more prevalent because these two modalities can provide complementary information to have the benefit of high temporal resolution and spacial resolution. In this case, gradient artifacts caused by a magnetic field is introduced into EEG recording. The requirement of gradient artifacts removal emerges and corresponding methods are proposed to remove such artifacts from EEG signal \cite{li2017unified, allen2000method, niazy2005removal}. Let us consider another situation that artifacts cannot be eliminated by routine methods. A new method (e.g., data completion method \cite{li2017new}) is required to tackle this problem. Therefore, the development of artifacts removal methods is not enough to tackle all cases. Efforts are continuing to be invested in the research of artifacts removal. We think that artifacts removal methods will be tailored to orient specific situations in the future, rather than for a general purpose. For example, a customized method will be developed to remove artifacts stemming from a particular movement and source (e.g., muscular contraction of the neck).

Currently, almost all experiments for cognitive state research are designed based on rational assumptions at a well-controlled environment (e.g., laboratory). Indeed, this is a feasible and effective way to probe into human cognitive states. However, can the assumptions be always true and accurate? A higher difficulty task might not always induce a heavier workload compared to a lower difficulty task. Possibly, the difference induced by two different difficulty levels of tasks is not distinct and is not enough to lead to differential characteristics in recorded signals (data). This assumption issue is more serious in the emotion studies. Images or videos with emotional contents are usually utilized to induce emotions of participants. It is assumed that emotion presented in an image or video is exactly induced in participants. This does not seem rigorous. The only thing guaranteed is that participants perceive different emotions presented in stimuli of images or videos. Participants might not be in the emotion as shown in the stimulus presented to them. An additional measure is necessary to determine whether the emotion shown in the stimulus is successfully induced. In addition, the scenario set up at a laboratory more or less differs from a scenario in daily life. Therefore, induced cognitive states at the laboratory are not the same as that appear in daily life. In the future, a more realistic scenario is required to investigate cognitive states so that mechanisms revealed in such a situation could be closer to ones they should be. For example, to explore the effect of walking on the human brain, overground walking is better than treadmill-based walking as overground walking is more naturalistic and the same to walking in daily life \cite{li2016robotic, li2018unilateral}. Similarly, cognitive states should be induced in a realistic scenario. For instance, neurophysiological signals and video are simultaneously recorded in a constraint-free conversation. Participants are not asked to generate a particular emotion at a particular time point. Instead, they are free to generate emotion in accordance to their feeling during the conversation. Emotion appears in the conversation is located based on the video recording and the corresponding neurophysiological signals during the appearance of the emotion can then be analyzed. Undoubtedly, such a realistic scenario setting dramatically increases setup difficulty and requires more effort in the following data processing. It might also enhance difficulty to control confounding factors. However, all extra costs are worthwhile in order to really understand the mechanisms of cognitive states.

 Among cognitive states discussed in this chapter, fatigue is relatively extensively investigated from the perspective of brain connectivity, ranging from laboratory-based study \cite{Harvy2018, Dimitrakopoulos2017a} to realistic study \cite{Wang2015a}. For the others, more studies are required to understand them further from a brain connectivity perspective. Using neurophysiological signals, human intentions can be decoded to establish a brain-computer interface for controlling external assistant devices \cite{Li2013, Muller-Putz2005, Pfurtscheller2003} or interacting with outside world \cite{Li2013a, Schalk2004, Gao2003}. The analogous systems for monitoring cognitive states will be more prevalent than ever. In order to facilitate practical usage, systems for monitoring cognitive states will be further miniaturized to be more portable and more comfortable to users. We have seen this trend in EEG recording devices. Dry electrodes have been adopted in the EEG recording system to reduce the preparation time before usage. The wireless technique has been applied for data transmission so that data cables, which restrains the mobile range of EEG recording, are not necessary anymore. Moreover, smart devices that have been used in daily life will be more deeply integrated into the monitoring systems of cognitive states. For example, we have developed a driving fatigue monitoring system, where a smartphone is integrated into the system to show fatigue level in real-time. In the future, appliances will be connected to the monitoring systems to better serve human by mutually sharing information between them. For example, the air conditioner automatically adjusts room temperature according to the monitoring data of human states. This can be seen in the near future because all the required techniques are ready. The new generation of communication technique (i.e., 5G) has been being used to largely enhance communication bandwidth, which makes it possible to have many things communicated with each other at the same time. By then, comprehensive information including cognitive states is collected from human and transferred to a clinician for cognitive assessment. As artificial intelligence is rapidly advanced, more and more work that has to be done by a clinician before will be taken over by artificial intelligence. With the benefit of super-speed communication technique, heavy computation demanded by artificial intelligence implementation can be shifted to a remote computational server. Such remote access to the computational resource is now available. Amazon web services (AWS) is one of the providers for the service. We believe that the use of the remote computational resource through the internet will gradually replace separate local computational workstations. Measured data are uploaded to a remote server for processing and analysing. The results are sent back to users for their review and record.

With more mechanisms relevant to cognitive states were revealed and the alterations of cognitive states were more deeply understood, a straightforward question appears in mind: is it possible to delay adverse changes in cognitive states and even to eliminate the adverse effect? A hint of possibility is given by the study of vigilance enhancement using visual challenge strategy \cite{Bodala2016}. Participants were requested to detect intruders in a simulated factory environment under two conditions with and without the presence of visual challenge (a rain obscuring the surveillance scene). They found that the visual challenge results in vigilance enhancement. Another study using a haptic stimulus showed that the haptic stimulus could help maintain vigilance \cite{Abbasi2017}. As we know, rest can help recover from fatigue and make the brain refreshed. This viewpoint was supported in terms of brain connectivity according to a study stated that a short break (rest) gave rise to a positive effect on preventing the changes of connectivity pattern towards fatigue \cite{Li2016}. These studies are still at very preliminary stage to address the mitigation or prevention of adverse changes of cognitive states, but the feasibility of cognitive modulation at some cases has been shown. In the future, a variety of manners other than those reported in the above studies will be investigated. This cognitive modulation will be integrated with a cognitive monitoring system so that an unwanted cognitive state could be detected and changed in real-time.

  \begin{figure}[!b]
 \centering
 \includegraphics[width=0.9\textwidth]{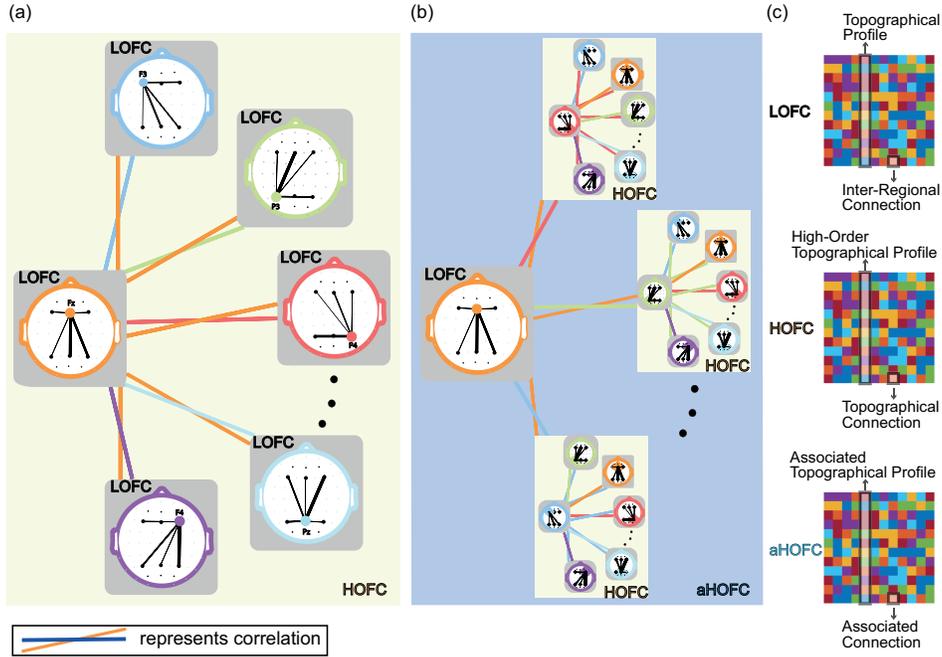}
 \caption{(a) High-order functional connectivity (HOFC) is obtained by correlation between topographical profiles, which are derived from connectivity strength estimation using any method presented in Sect. \ref{1.2}. These connectivity strength estimations represent inter-regional connections, which constitute low-order functional connectivity (LOFC). (b) Associated high-order functional connectivity (aHOFC) is a measure assessing similarity between the topographical profile and the high-order topographical profile. (c) Illustration of connections and profiles of the LOFC, HOFC, and aHOFC. Reprinted from \cite{li2020brain} with permission of IEEE.}
 \label{HOFCDisplay}
 \end{figure}  
                    
Machine learning methods have been successfully applied to decode brain activity to recognize cognitive states (e.g., fatigue level). Up to now, only classical and simple classifiers have been utilized to distinguish different levels of cognitive states based on connectivity features. More sophisticated classification methods are required to improve classification performance. In the past decade, deep learning has attracted much attention and has been very successful in image captioning, segmentation and recognition \cite{badrinarayanan2017segnet, he2016deep, Karpathy2015, papandreou2015weakly, wan2014deep}. It also exhibits promising performance in EEG classification \cite{li2014deep, Goh2018, Li2015, Jirayucharoensak2014, Cecotti2011}. In the future, deep learning will also be used to classify connectivity features to infer cognitive states. Before that, efforts should be paid to propose a suitable model structure and parameter setting. Another point we want to mention is that cognitive state classification only focuses on within-category classification. This is, the classification aims to differentiate different levels of the same cognitive state (e.g., fatigue levels), rather than to distinguish one cognitive state from another (e.g., fatigue versus workload). In the future, the cross-category classification will be performed. Ideally, a trained classifier can recognize different levels of diverse cognitive states. Furthermore, all of these levels of cognitive states can be classified in real-time. It is worth noting that low-order functional connectivity is currently used for classifying cognitive states. As shown in the latest study, high-order functional connectivity provides additional information characterizing mental fatigue \cite{Harvy2019} (see Fig. \ref{HOFCDisplay} for the illustration. High-order connectivity metrics can be extended as dynamic high-order connectivity metrics by incorporating temporal information over time.). These features of the high-order functional connectivity can be fused into a feature pool for performance improvement in the classification of cognitive states.

\section{Conclusion}
\label{Conclusion}
In this chapter, we reviewed the researches of cognitive states based on brain connectivity. An overview was first given to depict what was included in this chapter and typical signal processing for the analysis and classification of cognitive states. We then described the approaches which have been utilized to estimate connectivity strength and graph theoretical metrics for characterizing the attributes of connectivity networks. Subsequently, we summarised the findings of the data analysis and performances of the classification for mental fatigue, mental workload, vigilance, and emotion. Finally, we discussed existing issues and limitations and gave considerations about future directions in this research topic. Overall, mental fatigue was more extensively investigated than the other three cognitive states. Although significant progress was made in the research of cognitive states based on brain connectivity, further efforts are required to address the issues and limitations existing in the current research of cognitive states. According to the achieved studies, it is undoubted that brain connectivity is one of the promising perspectives to understand cognitive states.     
        
\bibliography{references}

\printindex
\end{document}